# A Probabilistic Embedding Clustering Method for Urban Structure Detection

Xin Lin[a,b], Haifeng Li[b*], Yan Zhang[b], Lei Gao[b], Ling Zhao[b], Min Deng[b]

[a] School of Civil Engineering, Central South University, Changsha, China
[b] School of Geosciences and info-physics, Central South University, Changsha, China

**Commission IV, WG IV/3**

**ABSTRACT:**

Urban structure detection is a basic task in urban geography. Clustering is a core technology to detect the patterns of urban spatial structure, urban functional region, and so on. In big data era, diverse urban sensing datasets recording information like human behaviour and human social activity, suffer from complexity in high dimension and high noise. And unfortunately, the state-of-the-art clustering methods does not handle the problem with high dimension and high noise issues concurrently. In this paper, a probabilistic embedding clustering method is proposed. Firstly, we come up with a Probabilistic Embedding Model (PEM) to find latent features from high dimensional urban sensing data by "learning" via probabilistic model. By latent features, we could catch essential features hidden in high dimensional data known as patterns; with the probabilistic model, we can also reduce uncertainty caused by high noise. Secondly, through tuning the parameters, our model could discover two kinds of urban structure, the homophily and structural equivalence, which means communities with intensive interaction or in the same roles in urban structure. We evaluated the performance of our model by conducting experiments on real-world data and experiments with real data in Shanghai (China) proved that our method could discover two kinds of urban structure, the homophily and structural equivalence, which means clustering community with intensive interaction or under the same roles in urban space.

## 1. INTRODUCTION

Urban structure detection is a basic task in urban geography, beneficial to urban planning, government management and so on (Bourne, 1963). In big data era, there are diverse urban sensing data to describe the city on different preservatives. Those data help us analysis and understand the process of urban structure in some aspects, such as urban functional regions (Yuan et al., 2012), socioeconomic environment (Liu et al., 2015), and sentiment computing (Xue et al., 2014). How to uncover the urban structures and discover patterns behind massive human behave data in unbans is vital. Clustering is one of core technique which utilizes the similarity between samples in a dense and short representation.

From a large body of clustering techniques (spectral clustering method (Luxburg, 2007; Ng et al., 2001), hierarchical clustering method (Rokach and Maimon, 2005), Fuzzy Clustering (Bezdek et al., 1984; Gustafson and Kessel, 1978), self-organization mapping (Kohonen and Honkela, 2007), etc.), it derives many relative works in urban structure exploration. For example, spectral clustering has already been used with success in urban land segmentation with social media data (Frias-Martinez and Frias-Martinez, 2014; Noulas et al., 2011); Hierarchical clustering method has been adopted to model urban structure with commuting flow data (Fusco and Caglioni, 2011); Google word2vec model also has been utilized to discover spatial distribution of urban land use with the data of points of interest (POI) (Yao et al., 2016).

However, some of the state-of-the-art clustering methods are expensive for large real-world networks since they involve eigendecomposition of the appropriate data matrix. Secondly, most of the methods rely on a rigid notion of a network neighbourhood, which fails to allow a flexible algorithm for a network mixed by complex structures (Tang and Liu, 2011). Moreover, the noise in real-world data calls for a robust algorithm. Consequently, facing the complexity in high-dimensional data, the key problem is how to find a compact and expressive feature, i.e. an embedding representation in low dimension feature space from a high-dimensional data space. And the approach needs to be flexible and robust.

In this paper, we proposed a probabilistic embedding clustering method to meet the need. Our contributions are listed as following.

Firstly, we proposed a Probabilistic Embedding Model (PEM) to learn embedded representation of high dimensional urban sensing data. With the probabilistic model, we can reduce uncertainty caused by high noise and high dimension. Then, by the representations, we do clustering and could catch essential features hidden in the data known as patterns of urban structure.

Secondly, our model could discover two kinds of urban structures, the homophily (Fortunato and Santo, 2010; Hoff et al., 2002) and structural equivalence (Henderson et al., 2012). The homophily contains nodes with intensive interaction or belonging to similar communities, while structural equivalence includes nodes having the same roles in urban structure.

The rest of the paper is structured as follows. We present the details of our method in Section 2. In Section 3, we evaluate the clustering performance of our method by conducting experiments on real-world data, including parameter sensitivity and perturbation analysis. In Section 4, we carry on a case study in Shanghai as an example for the detection urban structure. In Section 5, we conclude our work and illuminated some directions for future works.

* Corresponding author: lihaifeng@csu.edu.cn

## 2. PROBABILISTIC EMBEDDING CLUSTERING (PEC) METHOD

In this section, we will present the details of our method, the Probabilistic Embedding Clustering (PEC) Method. PEC is based on the algorithm, Node2vec (Grover and Leskovec, 2016), which is initiated to learn continuous feature representations for nodes in networks. By extending the meaning of networks (directed and weighted) and combining with a clustering method, we have developed the node embedding method into a general embedding-clustering method without changing its basic idea. There are three components of PEC method: (1) Construction of Space Relation Graph (SRG), (2) Probabilistic Embedding Model (PEM), (3) Clustering. We would discuss each of these in detail in the remaining part.

### 2.1 Construction of Space Relation Graph (SRG)

When recording the information of a city, we are used to divide the city into spatial parcels and the data stream is transformed to a set of arrays which describe the features of each parcel (Figure 1(a)). As known, the features could be intrinsic properties, like the size, population, number of Point of Interests (POIs) and so on (Figure 1(b)), but also the interaction with other parcels, like orientation-destination flow (Figure 1(c)). The set of arrays is called Urban Data Matrix. With such a matrix, we aim to find urban structure.

In essence, the goal requires us to summarize the similarity or dissimilarity among different parcels according to their features. Thus, we proposed the Space Relation Graph (SRG). In the graph, each node presents the parcel of the city and each edge denotes the relationship between parcels. As the features, the relationship could represent for the similarity of intrinsic properties among parcels (SRG-I), but also the degree of interaction among parcels (SRG-II). The only restriction is that the graph ought to be undirected.

For SRG-I, we measure the similarity by pairwise distance or other similarity function, such as Euclidean distance, cosine distance or Gaussian similarity function. While for SRG-II, since it is naturally a network, we simply normalize the pairwise interaction volume as the weight of edges.

### 2.2 Probabilistic Embedding Model (PEM) & Hyperparameters

After obtaining SRG, we learn a probabilistic embedding representation (PER) for each node based on Node2vec(Grover and Leskovec, 2016) which is to maximize the likelihood of preserving network neighbourhoods of nodes and learn a mapping of nodes to a low-dimensional space for obtaining feature representations. PER is such a feature representation existing in a low-dimensional space for each node.

The basic idea of PEM is utilizing the relationship between a node and the neighbours of it to represent the node itself. A fundamental hypothesis is the representation of a node in a network should be decided by its neighbourhood. The idea is similar to the Tobler's First Law, *"Everything is related to*

*everything else, but near things are more related than distant things"* (Tobler, 1970). To determine the neighborhood, PEM uses a biased random walk of $2^{nd}$ order Markovian. It defines two types of search strategies, the breadth-first search (BFS) (Kurant et al., 2010; Skiena, 2008) and the depth-first search (DFS) (Tarjan, 1972). BFS samples immediate neighbors of the source node while DFS samples neighbors sequentially at increasing distances from the source node. The strategies are controlled by hyperparameters $p$ and $q$.

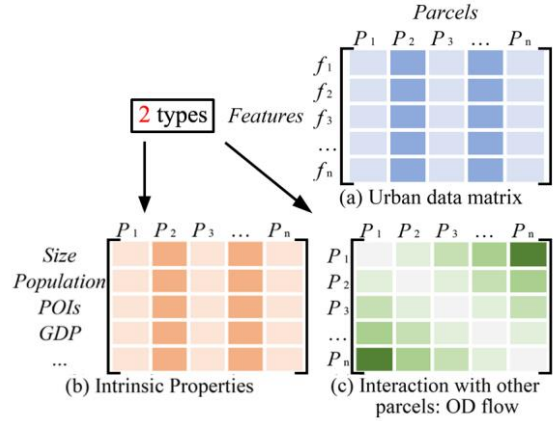

Figure 1. Urban data matrix (a): information of parcels is stored in arrays and thus the urban data matrix is formed. There are 2 types of urban data matrix. One is to describe intrinsic properties of parcels (b) and the other one is to describe the interaction among parcels (c).

Given a source node t, and that the walk just sampled the node v as the neighbour and is now carrying out its next step, the biased random walk works as Figure 2 and it shows how hyperparameters $p$ and $q$ drive the walk.

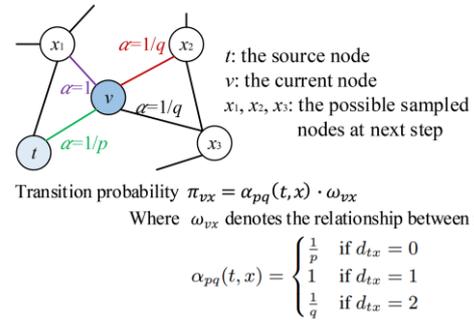

$t$: the source node
$v$: the current node
$x_1, x_2, x_3$: the possible sampled nodes at next step

Transition probability $\pi_{vx} = \alpha_{pq}(t,x) \cdot \omega_{vx}$
Where $\omega_{vx}$ denotes the relationship between nodes from SGR

$$\alpha_{pq}(t,x) = \begin{cases} \frac{1}{p} & \text{if } d_{tx}=0 \\ 1 & \text{if } d_{tx}=1 \\ \frac{1}{q} & \text{if } d_{tx}=2 \end{cases}$$

Figure 2. Illustration of the biased 2nd random walk procedure in node2vec: It shows that how hyper-parameters $p$ and $q$ guide the walk between BFS and DFS.

Besides the hyperparameters, there are also other global parameters involved in the random walk and effective in the clustering results (Table 1). Unfortunately, there is no systematic study investigating the effects of parameters and hyperparameters on clustering and coming up with well-justified rules of thumb. As a general recommendation, we suggest to set $p$ to a lower value and $q$ to a higher value for BFS while set $p$ to a higher value and $q$ to a lower value for DFS. If you want to balance the walk between BFS and DFS, a low $q$ and a low $p$ (a high $q$ and a high $p$) are recommended. When ($p$, $q$) = 1, the random walk is just driven by the weights in SRG. Other parameters are suggested to be set higher for a more accurate clustering result. It would saturate around some values definitely but no firm theoretic ground is proposed towards various networks. All these recommendations would be examined in Section 3.

| Parameters | Definition |
|---|---|
| $d$ | Number of dimensions of the embedding vector for feature representation. |
| $l$ | Walk length which determines the length of sampling from a source node. |
| $r$ | Number of walks which denotes the times of sampling from a source node. |
| $k$ | Neighbourhood size which restricts the size of neighbourhood. |

Table 1. Global parameters involved in the biased random walk: Setting them to higher values would help obtain a more accurate clustering result.

### 2.3 Clustering

We use k-means algorithms on PER to get final results. Speaking from experience, when adopting BFS ($p \leqslant 1$, $q > 1$), we could uncover the community where the subjects have intensive interaction, named as homophily. While with DFS ($p \geqslant 1$, $q < 1$), we could divide the subjects into a hierarchical pattern where the subjects play the same role within a grade, named as structural equivalence. The two types of structure make great sense in urban space: The homophily could help describe the phenomenon of spatial separation and the structural equivalence is conducive to find the urban hierarchy.

Though the search strategy and other parameters influences the detection of urban structure, the network itself restricts the detection. For example, for SRG with actual interaction, we are likely to find the homophily and structural equivalence. For SRG with virtual similarity between the property like functions of places, we utilize BFS mainly to explore the urban functional regions.

To determine the number of clusters, we use several clustering criteria: Davies-Bouldin index (Davies and Bouldin, 1979), Dunn index (Dunn, 1973), Silhouette index (Rousseeuw, 1987) and so on. Of course, other clustering criteria is welcomed as well. In certain cases, the number of clusters could be also decided by priori knowledge.

## 3. EVALUATION

In this section, we mainly evaluate the performance of our method on a real-world network, Metro Line Network in Shanghai (China), including parameter sensitivity, perturbation analysis and so on.

### 3.1 Experimental Setup

**Dataset**: Line 1 to 11 of Shanghai metro network are used in this section and we transformed the metro line network in Shanghai to an urban data matrix in this way: if the two subway stations $s_i$ and $s_j$ are connected directly, the pairwise distance is denoted as 1. Otherwise, we regard the distance is infinitely big. Thus, the connection among subway stations is used to represent the stations themselves. Here, SRG is equal to the data matrix.

**3.1.1 Dataset**: Line 1 to 11 of Shanghai metro network are used in this section and we transformed the metro line network in Shanghai to an urban data matrix in this way: if the two subway stations $s_i$ and $s_j$ are connected directly, the pairwise distance is denoted as 1. Otherwise, we regard the distance is infinitely big. Thus, the connection among subway stations is used to represent the stations themselves. Here, SRG is equal to the data matrix.

**3.1.2 Ground-truth**: The metro line network is mixed with homophily and structural equivalence. With using SRG of metro line network in the way mentioned above, it is much easier to discover the structural equivalence. The network only has a hierarchy of two levels: transfer station or not. Thus, the number of clusters ($n$) are taken as 2 and the corresponding ground-truth have two kinds of labels marking transfer station or not. For another thing, the homophily definitely means clusters with subway stations on the same line. Thus, the number of clusters are taken as 11 and the corresponding ground-truth has 11 unique labels marking which line the subway station is on. With using this SRG, it needs to catch information at a larger neighbourhood for obtaining structure of homophily.

**3.1.3 Comparison groups**: Our experiments evaluate the embedding vectors obtained by PEM on clustering. For the task, we compare the performance of PEM with that of the following clustering algorithms:

Spectral clustering (SC) (Frias-Martinez and Frias-Martinez, 2014; Luxburg, 2007): SC is a clustering method which also construct a similarity graph at first. However, it is a naturally factorization approach where we take the top $d$ eigenvectors of the normalized Laplacian matrix of the graph as the embedding vector for nodes. And then it uses k-means to do clustering.

Hierarchical clustering analysis (HCA) (Fusco and Caglioni, 2011; Rokach and Maimon, 2005): HCA is a clustering method which seeks to build a hierarchy of clusters. There are 2 types of strategies: Agglomerative (a bottom-up approach merging observations from each one) and Divisive (a top-down approach dividing the observations from one cluster). Here, we adopt Agglomerative HCA.

All the experiments would be repeated 20 times (if possible) and the average is taken as the final results.

### 3.2 Hyperparameters

We regard that BFS and DFS are guided by hyperparameters $p$ and $q$ and thus influence whether the detected result of urban structure is homophily or structural equivalence. We now empirically show this phenomenon. The other parameters are set as $d$=5, $l$=10, $r$=10 and $k$=5.

As mentioned in Section 3.1.2, we could deduce that adopting BFS helps discover the structural equivalence (when n=2) while using DFS helps find the homophily (when n=11) in this network. Figure 3 shows that when the search strategy is more inclined to BFS ($p$<1 or $q$ >1), Macro-F1 scores of structural equivalence get higher and when the search strategy is more inclined to DFS ($p$>1 or $q$<1), Macro-F1 of homophily get higher. Additionally, with most settings of hyperparameters, PEM outperforms SC and HCA in both cases. Thus, it proves that PEM is flexible, which is able to discover different kinds of structure through tuning hyperparameters. And it is also accurate when the values of $p$ and $q$ are proper.

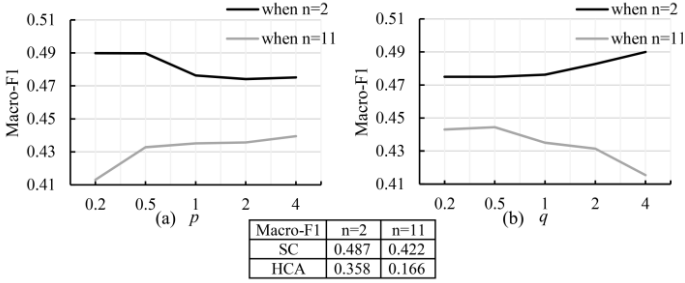

| Macro-F1 | n=2 | n=11 |
|---|---|---|
| SC | 0.487 | 0.422 |
| HCA | 0.358 | 0.166 |

Figure 3. Changes of Macro-F1 versus hyperparameters $p$ and $q$: when the search strategy is more inclined to BFS (p<1 or q >1), Macro-F1 scores of structural equivalence get higher and when the search strategy is more inclined to DFS (p>1 or q<1), Macro-F1 of homophily get higher. Additionally, with most settings of hyperparameters, PEM outperforms SC and HCA in both cases.

### 3.3 Parameter Sensitivity

PEM involves a set of parameters and we examine the recommendations about how to choose the parameters in Section 2.2. Here, the basic parameters are set as $d$=5, $l$=10, $r$=10, $k$=5 and the hyperparameters are set as $p$=4, $q$=1. The influences of parameters on the detection of homophily and structural equivalence in this network are shown as Figure 4.

As seen in Figure 4, the overall effect of parameters is positive: Macro-F1 scores are on the rise despite the fluctuations. It is also could be seen that the parameters have little effect on the results when n=2. As mentioned above, when $p$=4 and $q$=1, it's DFS guiding the random walk and we are likely to discover structure of homophily in this network. Thus, under the significant influence of hyperparameters, it is reasonable that Macro-F1 scores are not hardly affected by the parameters when n=2 while Macro-F1 scores rise obviously. Among the parameters, $k$ has a significant influence on the clustering, and $d$ as well as $r$ follows. The influence of $l$ is limited.

### 3.4 Perturbation Analysis

Since many real-world datasets co-exists with uncertain noise, we perform a perturbation study where we added noise data. The distribution of noise obeys Gaussian distribution N(0,$\sigma^2$) ($\sigma$ $\in$ (0.2,0.5,1,2,4)) and Poisson distribution P($\lambda$) ($\lambda$ $\in$ (1,2,4,8,16)) respectively and the value of created noise matrices is processed into [0, 1]. The experimental results are shown as following. It could be seen that Macro-F1 scores change little whatever the distribution of noise is. Thus, PEM is proved to be able to handle the problem with high noise.

## 4. CASE STUDY: URBAN STRUCTURE IN SHANGHAI USING TAXI TRAJECTORY DATA

### 4.1 Experimental Setup

*The studied area*: We take the central area within the outer ring of Shanghai as the studied area and divide the area into 4422 uniform grids whose size is 500m×500m (Figure 6).

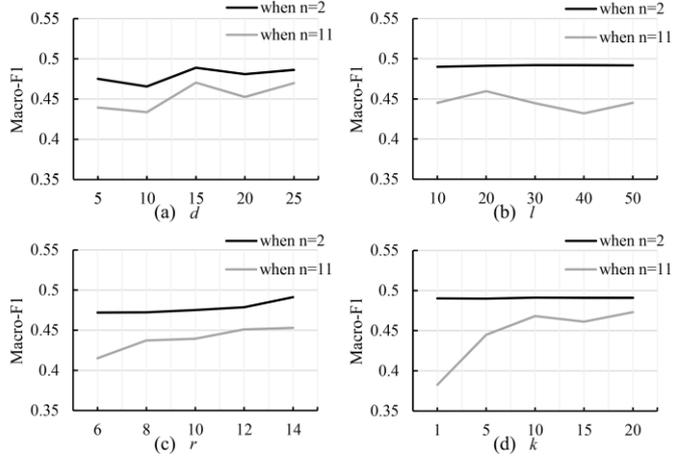

Figure 4. Parameter Sensitivity: the overall effect of parameters is positive: Macro-F1 scores are on the rise despite the fluctuations[1].

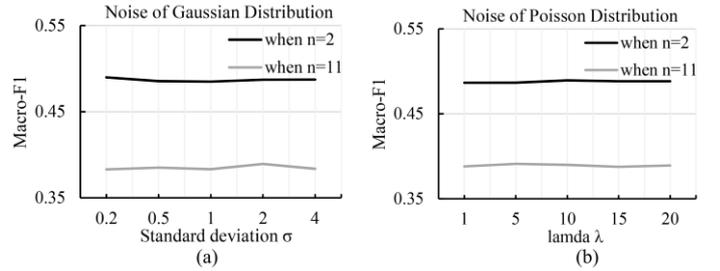

Figure 5. Perturbation analysis: The steady Macro-F1 results show that PEM could handle the problem with high noise

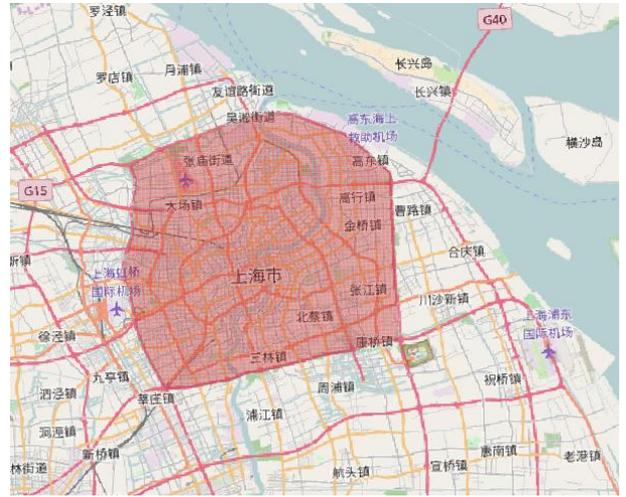

Figure 6. The studied area: We take the central area within the outer ring of Shanghai as the studied area and divide the area into 4422 uniform grids whose size is 500m×500m.

#### 4.1.1 Dataset: We use the dataset containing trajectories of over 160 million records created by more than 4,000 taxis from December 2nd to December 5th in 2013 (a four-day pattern). Derived from the trajectory datasets, we obtain the orientation-destination matrix about 4422 grids.

---

[1] Since the accurate values of experimental results are not repeatable, the Macro-F1 scores have a little difference. But the trends reflected by the experiments are provable.

**4.1.2 Experimental settings**: We set hyperparameters and parameters as $p=1$, $q=1$, $d=64$, $l=80$, $r=10$ and $k=5$. The number of clusters is determined to be 5 according to the fast unfolding algorithm.

## 4.2 Result and Discussion

As $p=1$ and $q=1$, the random walk is balanced between BFS and DFS. The detection result (Figure 7) of urban structure would thus reflect the interaction between regions faithfully. As Figure 7(a) shows, the regions are cohesive and have clear boundary. Figure 7(b) shows that it almost confirms to the boundary of administrative districts. It could be seen that people seldom take taxi for a further place and the interaction frequency within a region are much higher than that between regions (Table 2). Consequently, PEM discovered the homophily structure of the studied area successfully.

| Freq. | Region 1 | Region 2 | Region 3 | Region 4 | Region 5 |
|---|---|---|---|---|---|
| Region 1 | **0.502** | 0.082 | 0.204 | 0.137 | 0.073 |
| Region 2 | 0.228 | **0.376** | 0.162 | 0.212 | 0.016 |
| Region 3 | 0.177 | 0.053 | **0.611** | 0.082 | 0.077 |
| Region 4 | 0.247 | 0.147 | 0.165 | **0.413** | 0.021 |
| Region 5 | 0.257 | 0.023 | 0.314 | 0.044 | **0.362** |

Table 2. The interaction frequency between regions: the interaction frequency within a region are much higher than that between regions.

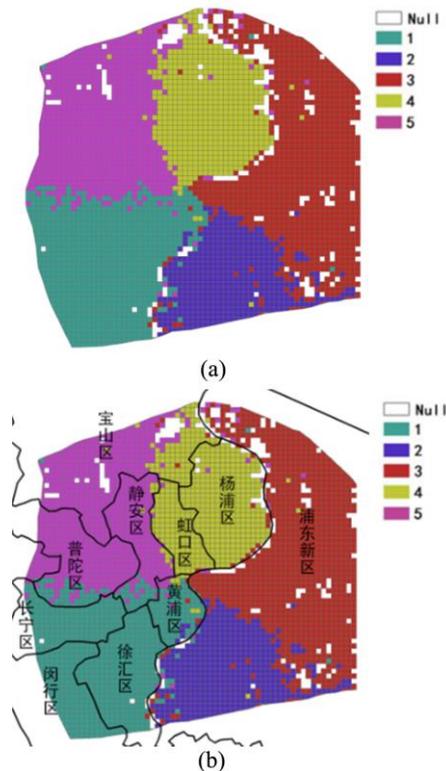

(a)

(b)

Figure 7. The detection result of urban structure in the studied area: (a) the visualization by ARCGIS; (b) the detection result compared with the administrative map: the regions are cohesive and have clear boundary which almost confirms to the boundary of administrative districts.

## 5. CONCLUSION AND FUTURE WORKS

Through the evaluation on the real-world network (the metro line network in Shanghai, China), PEM is flexible to explore urban structure through tuning hyperparameters and outperforms Spectral Clustering and Hierarchical Clustering Analysis with proper settings of parameters. In usual, when $p$ is lower or $q$ is higher, it is inclined to adopt BFS. In contrast, it is inclined to adopt DFS. In different networks, BFS (or DFS) would find different type of structure. While in general, the influence of parameters (dimension $d$, walk of length $l$, number of walks $r$ and context size $k$) have a positive influence on the accuracy. Additionally, the perturbation analysis verifies that PEM is able to learn feature presentations for nodes and do clustering under high dimension and high noise.

Based on the study, with the taxi dataset in Shanghai, we utilized PEM to explore the structure of region distributions successfully. However, to improve the theory and method, there is still a long way to go.

In the future work, we would add more experiments with different urban sensing data and other 2vec models and fully utilize the flexibility of PEM to explore urban structure. And the systematic study of setting parameters and link with other theories would be extended.

**ACKNOWLEDGEMENTS**

This research has also been supported by National Natural Science Foundation of China (41571397 and 41501442) and Natural Science Foundation of Hunan Province (2016JJ3144) .

This research has also been supported in part by the Open Research Fund Program of Shenzhen Key Laboratory of Spatial Smart Sensing and Services (Shenzhen University), and the Scientific Research Foundation for the Returned Overseas Chinese Scholars, State Education Ministry (50-20150618).